\documentclass[prb,showpacs,twocolumn,byrevtex]{revtex4}
\usepackage{dcolumn}
\usepackage{graphicx}
\usepackage{float}
\begin{document}

\bibliographystyle{apsrev}
%bibliographystyle{prsty}

\preprint{Draft, version 2.1 (not for distribution)}
%
% The title and the list of authors
%
\title{Coherence, incoherence and scaling along the \boldmath $c$ axis
 of YBa$_2$Cu$_3$O$_{6+x}$ \unboldmath}
\author{C. C. Homes}
\email{homes@bnl.gov}
\author{S. V. Dordevic}
\affiliation{Department of Physics, Brookhaven National Laboratory, Upton, New
York 11973}%
\author{D. A. Bonn}
\author{Ruixing Liang}
\author{W. N. Hardy}
\affiliation{Department of Physics and Astronomy, University of British
Columbia, Vancouver, B.C. V6T 2A6, Canada}%
\author{T. Timusk}
\affiliation{Department of Physics and Astronomy, McMaster University,
Hamilton, Ontario  L8S 4M1, Canada}%
\date{\today}

%
% The abstract goes here
%
\begin{abstract}
The optical properties of single crystals of YBa$_2$Cu$_3$O$_{6+x}$ have been
examined along the {\it c} axis above and below the critical temperature
($T_c$) for a wide range of oxygen dopings.  The temperature dependence of the
optically-determined value of the dc conductivity ($\sigma_{dc}$) in the normal
state suggests a crossover from incoherent (hopping-type) transport at lower
oxygen dopings ($x \lesssim 0.9$) to more coherent anisotropic
three-dimensional behavior in the overdoped ($x\approx 0.99$) material at
temperatures close to $T_c$. The assumption that superconductivity occurs along
the {\it c} axis through the Josephson effect yields a scaling relation between
the strength of the superconducting condensate ($\rho_{s,c}$, a measure of the
number of superconducting carriers), the critical temperature, and the
normal-state {\it c}-axis value for $\sigma_{dc}$ just above $T_c$; $\rho_{s,c}
\propto \sigma_{dc}\,T_c$. This scaling relation is observed along the {\it c}
axis for all oxygen dopings, as well as several other cuprate materials.
However, the agreement with the Josephson coupling model does not necessarily
imply incoherent transport, suggesting that these materials may indeed be
tending towards coherent behavior at the higher oxygen dopings.
\end{abstract}
%
%  PACS numbers
%  63.20.-e  Phonons in crystal lattices
%  77.22.Ch  Permittivity (dielectric function)
%  78.30.-j  Infrared and Raman spectra
%
\maketitle
\vspace*{-0.6cm}
%
% The main body of the text
%
% Introduction
%
\section{Introduction}
The cuprate-based high-temperature superconductors all share the common feature
that the superconductivity is thought to originate within the highly-conducting
copper-oxygen planes.  The conductivity perpendicular to the planes along the
{\it c} axis is much poorer and is in fact activated in many cuprates; the
resulting transport is due to hopping along this direction and the large
anisotropy in the resistivity results in the two-dimensional (2D) nature of
these systems.\cite{cooper94}  The importance of the 2D character of these
materials as a necessary prerequisite for superconductivity has been
discussed,\cite{anderson98} and it has recently been suggested that upon
entering the superconducting state a dimensional crossover from two to three
dimensions occurs.\cite{valla02,menzel02} Transport measurements on a number of
cuprate systems suggest that such a dimensional crossover may occur in the
normal state in response to carrier doping in the copper-oxygen
planes.\cite{ito91,nakamura93,kao93,hussey97,schneider02, schneider04}  A
convenient system to examine is YBa$_2$Cu$_3$O$_{6+x}$ (YBCO), where the oxygen
doping determines not only the in-plane carrier concentration, but also the
nature of the {\it c}-axis transport. This material is one of the most
thoroughly studied, and there have been numerous reports of the response of the
physical properties to changes in oxygen doping, including
transport\cite{ito91,takenaka94} and optical
techniques,\cite{cooper93,homes93a,homes95a,homes95b,homes98,schutzmann94,
schutzmann95b,tajima97} to name but a few.

%
% Low dopings...
%
In YBCO oxygen dopings below the optimal value $(x=0.95)$ for the critical
temperature $T_c$, the dc resistivity along the {\it c} axis is $\rho_c \propto
T^\alpha$, where $\alpha \lesssim -1$, indicative of activated behavior. In
these underdoped materials, the anisotropy between the {\it a-b} planes and the
{\it c} axis, as gauged by the resistivity, is quite large: $\rho_c/\rho_a
\gtrsim 65$ at room temperature, and increases rapidly with decreasing
temperature to $\rho_c/\rho_a \geq 3000$ ($x\lesssim 0.70$) just above $T_c$
(Ref.~\onlinecite{takenaka94}). For these reasons, transport along the {\it c}
axis in the normal state in the underdoped materials is governed by hopping and
is considered to be incoherent; below $T_c$ superconductivity normal to the
planes is thought to involve Josephson
coupling.\cite{shibauchi94,basov94,radtke95a}
%
% Discuss the properties of the overdoped systems.
%
However, as the material becomes nearly stoichiometric, or ``overdoped'' ($x
\approx 0.99$), the {\it c}-axis properties change dramatically. Unlike in the
underdoped systems, the resistivity rises linearly with $T$, although with a
large, temperature-independent component. Furthermore, $\rho_c/ \rho_a \lesssim
30$ at room temperature, and is nearly temperature independent in the normal
state,\cite{takenaka94} characteristic of an anisotropic three-dimensional (3D)
metal.  This behavior is even more pronounced in the Ca-doped
material.\cite{bernhard99}  In addition, the anisotropic resistivity of the
double-chain material YBa$_2$Cu$_4$O$_8$ also suggests an
incoherent-to-coherent crossover in the out-of-plane behavior.\cite{hussey97}
This suggests that transport along the {\it c}-axis of YBCO is becoming more
coherent at high oxygen dopings, and that a crossover from 2D to 3D behavior
may occur; this could have important consequences for the nature of the
superconductivity.

%
% What we do in this work
%
In this work we examine the optical properties along the {\it c} axis of
YBa$_2$Cu$_3$O$_{6+x}$ for a wide range of oxygen dopings in the normal and
superconducting states.  The normal-state transport suggests that a dimensional
crossover may occur in the overdoped YBCO samples.  The strength of the
condensate along the {\it c} axis is consistent with the Josephson coupling
between the planes. However, because the Josephson effect may be observed for
both coherent and incoherent transport along the {\it c} axis (tunnel and
non-tunnel junctions), the condensate does not necessarily favor incoherent
transport in the overdoped material, where the normal-state properties suggest
the system is tending towards coherent behavior.

%
% Experimental section
%
\section{Experiment}
Large, twinned single crystals of YBCO were grown by a flux method in
yttria-stabilized zirconia crucibles\cite{liang92} and subsequently reannealed
to a variety of oxygen contents from $x=0.50\rightarrow 0.99$. The $T_c$'s are
sharply defined and vary over a large range, from a low of $T_c=53\,$K in the
most underdoped material ($x=0.50$) to a maximum of $T_c=93.2\,$K in the
optimally doped material ($x=0.95$).  In the most overdoped sample ($x\gtrsim
0.99$), $T_c$ is suppressed somewhat to $T_c\approx 90\,$K.  The reflectance
polarized along the {\it c} axis was measured over a wide frequency range and
at a variety of temperatures using an overfilling technique.\cite{homes93b}
The real part of the optical conductivity $\sigma_1(\omega)$ was determined
from a Kramers-Kronig analysis of the reflectance.

%
% Figure 1
%
\begin{figure}[t]
%\vspace*{-0.7cm}
%
% eprint
%
\centerline{\includegraphics[width=3.2in]{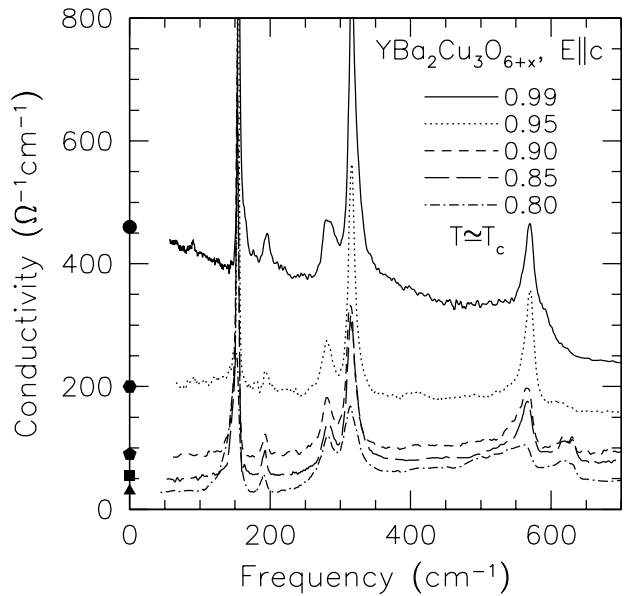}}%
%\vspace*{-0.1cm}%
%
% manuscript
%
%\centerline{\includegraphics[width=6.5in]{figure1.eps}}%
%\vspace*{-1.5cm}%
%
\caption{The real part of the optical conductivity for light polarized along
the {\it c} axis of YBa$_2$Cu$_3$O$_{6+x}$ at $T\gtrsim T_c$ for a variety of
oxygen dopings.  The extrapolated values of the dc conductivity $\sigma_{dc} =
\sigma_1(\omega \rightarrow 0)$ are shown as symbols.  The most heavily doped
regime ($x=0.99$) is representative of a weakly-metallic system and the
conductivity displays a Drude-like frequency dependence.  The lowest oxygen
concentration shown here ($x=0.80$) is in the pseudogap regime, with an
activated frequency response.}%
%\vspace*{-0.3cm}%
\label{fig:sigma}
\end{figure}

%
% Figure 2
%
\begin{figure}[t]
%\vspace*{-0.7cm}
%
% eprint
%
\centerline{\includegraphics[width=3.2in]{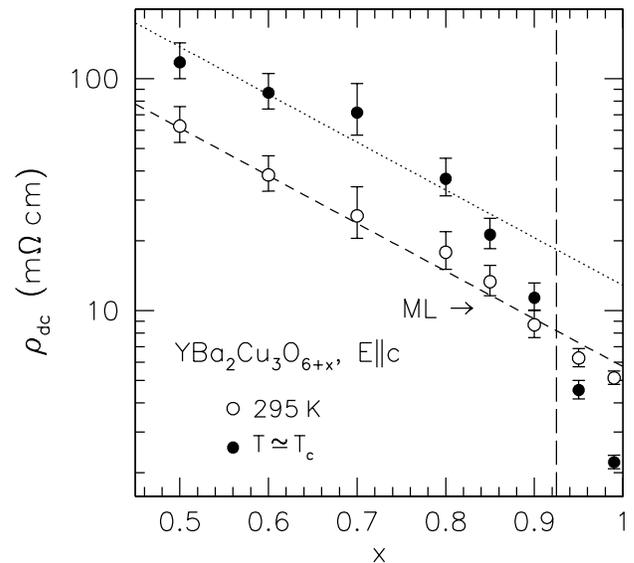}}%
%\vspace*{-0.1cm}%
%
% manuscript
%
%\centerline{\includegraphics[width=6.5in]{figure2.eps}}%
%\vspace*{-1.5cm}%
%
\caption{Resistivity $\rho_{dc} = 1/\sigma_1(\omega \rightarrow 0)$ for
YBa$_2$Cu$_3$O$_{6+x}$ along the {\it c} axis at 295~K and $T \gtrsim T_c$ as a
function of oxygen doping.  At room temperature, $\rho_{dc}$ increases exponentially
with decreasing oxygen doping, as indicated by the dashed line. However, for
$T\gtrsim T_c$, a linear fit applies only to the dopings between $0.50\leq x \leq
0.85$, or the underdoped region (dotted line); values for $x
> 0.9$ fall below this line.  This indicates that in the underdoped region
resistivity is increasing with decreasing temperature, while in the overdoped
region precisely the opposite behavior is observed at roughly the Mott minimum
limit for metallic conductivity (ML).}%
%\vspace*{-0.3cm}%
\label{fig:resis}
\end{figure}

%
% The normal state - transport properties
%
\section{Results and Discussion}
\subsection{Normal State}
The optical properties along the {\it c} axis have been thoroughly investigated
by us and others,\cite{cooper93,homes93a,homes95a, homes98,schutzmann94,
schutzmann95b,tajima97} and will be discussed only briefly. The optical
conductivity of YBCO is shown along the {\it c} axis in Fig.~\ref{fig:sigma}
for $T\gtrsim T_c$ at variety of oxygen dopings.  In the overdoped regime the
conductivity has a metallic temperature and frequency dependence.  However,
with decreasing doping a pseudogap develops, and a non-metallic activated
response is observed.
A convenient connection between the optical properties and transport is that
$\sigma_{dc}\equiv \sigma_1(\omega\rightarrow 0)$ as indicated in the plot and
listed in Table~I (for $T\gtrsim T_c$).  The extrapolated values for the
resistivity $\rho_{dc}$ along the {\it c} axis are shown in
Fig.~\ref{fig:resis} for a variety of oxygen dopings at room temperature and $T
\gtrsim T_c$.  At room temperature the resistivity increases exponentially with
decreasing doping, $\rho_c \propto \rho_0\, e^{-ax}$, throughout the entire
doping range.  At low temperature the resistivity is observed to increase with
decreasing temperature for $x \lesssim 0.85$, in agreement with the activated
response observed in transport; once again the resistivity is varying
exponentially with doping. However, for $x \gtrsim 0.9$ the resistivity
decreases dramatically with increasing doping and no longer follows the simple
exponential relation.  In addition, the temperature response of the resistivity
is now ``metallic'', particularly so for the overdoped sample.\cite{ito91} This
change in behavior also occurs close to the Mott maximum value for metallic
behavior (ML), estimated to be $\rho_c \approx 10$~m$\Omega\,$cm in these
materials.\cite{mott}
%
% Similar behavior has also been cited as evidence for a dimensionality crossover
% in La$_{2-x}$Sr$_x$CuO$_4$ in the ``overdoped'' regime ($x\gtrsim
% 0.34$).\cite{nakamura93,kao93,ito91}
%
% While the normal-state reflectance of the overdoped  material never develops a
% sharp plasma edge such as the one associated with the  although this may be due
% the fact that any such trend is interrupted
%
%
% optical properties of the overdoped system...
%
The overdoped material is clearly the most metallic, and displays a Drude-like
conductivity, an important result that was noted in earlier
studies.\cite{schutzmann94} The Drude model for the dielectric function
describes the properties of a simple metal quite well, $\tilde\epsilon(\omega)
= \epsilon_\infty - \omega_p^2/[\omega(\omega + i\Gamma)]$, where $\omega_p$ is
the classical plasma frequency, $\Gamma = 1/\tau$ is the scattering rate, and
$\epsilon_\infty$ is a high-frequency contribution.  The Drude conductivity is
$\sigma_1(\omega)= \sigma_{dc}/(1 + \omega^2\tau^2)$, which has the form of a
Lorentzian centered at zero frequency with a width at half-maximum of $1/\tau$.
For $T\gtrsim T_c$, the optical conductivity of the overdoped system shown in
Fig.~\ref{fig:sigma} does have a Drude-like frequency response,
$\sigma_1(\omega) \propto 1/\omega^2$. In addition $\sigma_{dc} \equiv
\sigma_1(\omega\rightarrow 0) \approx 450\,\Omega^{-1}\,{\rm cm}^{-1}$ for
$T\gtrsim T_c$ is well above the Mott minimum value for metallic conductivity
of $\approx 100$~$\Omega^{-1}\,{\rm cm}^{-1}$ in these materials.

%
% Results of a Drude fit...
%
Previous investigations of overdoped systems resulted in large values for both
the plasma frequency $\omega_{p,c}$ ($\gtrsim 4000\,$cm$^{-1}$) and the
scattering rate $1/\tau_c$ ($\gtrsim 1000\,{\rm cm}^{-1}$).\cite{cooper93,
schutzmann94} A principal objection to coherent transport along the {\it c}
axis has been the large values for $1/\tau_c$ which lead to mean free paths
that are substantially less than a lattice spacing.\cite{cooper94}
If the {\it c}-axis conductivity is modeled using a two-component approach
(Drude component to model free carriers, plus Lorentzian oscillators to
represent bound excitations), then it is possible to estimate the {\it c}-axis
plasma frequency and scattering rate.  This approach yields values of
$\omega_{p,c} \simeq 3200\,$cm$^{-1}$ and $1/\tau_c \simeq 380\,$cm$^{-1}$ at
100~K, smaller than previously-observed values.\cite{mir}  These results may be
compared to the values in the copper-oxygen planes along the {\it a} axis for
the overdoped material\cite{room} of $\omega_{p,a} \simeq 10\,000\,{\rm
cm}^{-1}$ and $1/\tau_a\simeq 120\,{\rm cm}^{-1}$. The resistivity anisotropy
is then expected to be $\rho_c/\rho_a = (\omega_{p,a}^2 \tau_a) /
(\omega_{p,c}^2 \tau_c) \gtrsim 30$, where $\omega_{p,a}^2 / \omega_{p,c}^2
\simeq 10$, which is in good agreement with band-structure
estimates.\cite{allen88} Furthermore, taking of $v_F=7\times 10^6\,$cm/s gives
a mean free path along the {\it c} axis ($l_c=v_F\tau_c$) of $l_c \simeq
60\,$\AA, which is more than five times the size of the unit cell, suggesting
the possibility of coherent (Bloch-Boltzmann) transport in the overdoped
material for $T\gtrsim T_c$ (Ref.~\onlinecite{schutzmann94}).  The increasingly
coherent transport along the {\it c} axis has also been discussed in relation
to the strength of the inelastic scattering in the copper-oxygen
planes.\cite{dordevic02}
%
% However, the uncertainties associated with $\omega_{p,c}$ and $1/\tau_c$ as
% determined from the two-component model in this study suggest, while st that
% any arguments for or against coherent transport based on the mean free path be
% treated with caution.

%
% Figure 3
%
\begin{figure}[t]
%\vspace*{-0.7cm}
%
% eprint
%
\centerline{\includegraphics[width=3.2in]{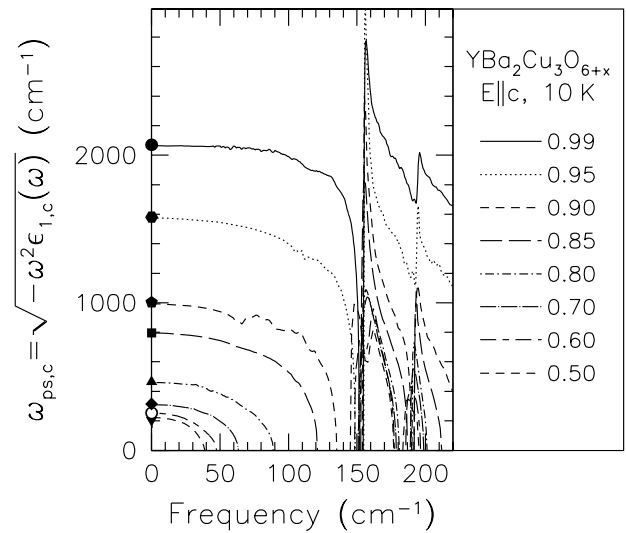}}%
%\vspace*{-0.1cm}%
%
% manuscript
%
%\centerline{\includegraphics[width=6.5in]{figure3.eps}}%
%\vspace*{-1.5cm}%
%
\caption{The doping-dependent behavior of $\sqrt{-\omega^2 \epsilon_{1,c}
(\omega)}$ for $T\ll T_c$.  In the $\omega \rightarrow 0$ limit this quantity
is the plasma frequency of the condensate along the {\it c} axis,
$\omega_{ps,c}$.  The low-frequency extrapolation employed in the
Kramers-Kronig analysis below $\simeq 40\,$cm$^{-1}$ are included as a guide to
the eye. The strength of the plasma frequency is decreasing dramatically with
decreasing oxygen doping (Table~I).}
%
%\vspace*{-0.3cm}%
\label{fig:rho}
\end{figure}

%
% Trends in the superconducting state
%
% Trends below T_c
%
\subsection{Superconducting State}
In YBCO the onset of superconductivity is accompanied by the dramatic formation
of a plasma edge in the reflectance along the {\it c} axis for all the oxygen
dopings studied.\cite{homes93a,homes95a} In the copper-oxide superconductors,
the order parameter is thought to originate within the planes, with bulk
superconductivity achieved through coherent pair tunneling between the planes
occurring from the Josephson effect.\cite{abrikosov98} In such a case the {\it
c}-axis penetration depth $\lambda_c$ is determined by the Josephson current
density $J_c$ and is $\lambda_c^2 = \hbar c^2 / 8\pi^2deJ_c \propto 1/dJ_c$,
where $d$ is the separation between the planes.\cite{bulaevskii73,lawrence71}
In the BCS theory, $J_c$ is related to the energy gap $\Delta(T)$ and the
tunneling resistance per unit area in the normal state $R_n$
by\cite{ambegaokar63}
\begin{equation}
  J_c =  {{\pi\Delta(T)}\over{2e R_n}}  \tanh
        \left[ {{\Delta(T)} \over {2k_B T}} \right].
\end{equation}
Adopting the BCS isotropic {\it s}-wave gap weak-coupling value of $\Delta(T\ll
T_c) \simeq 1.76\,k_B\,T_c$ and assuming that $R_n = \rho_{dc}\,d$, then $J_c
\propto T_c/R_n$ at low temperature, and $1/\lambda_c^2 \propto
\sigma_{dc}\,T_c$, where $\sigma_{dc}$ is the extrapolated dc conductivity
along the {\it c} axis in the normal state $(T\gtrsim
T_c)$.\cite{shibauchi94,basov94}  From $1/\lambda = 2\pi \omega_{ps}$, the
strength of the condensate is $\rho_s \equiv \omega_{ps}^2$, yielding
\begin{equation}
  \rho_{s,c} \simeq 65\,\sigma_{dc}\,T_c,
  \label{eq:jc}
\end{equation}
where the right and left hand side of the expression have units of cm$^{-2}$.
%
% Determination of the condensate
%
We note that several other workers have arrived at a similar relationship based
on different assumptions.\cite{smith92,chakravarty99}  If the clean limit is
assumed (all the normal-state carriers collapse into the condensate), then for
$T\ll T_c$ the response of the dielectric function is purely real,
$\tilde\epsilon(\omega) \equiv \epsilon_1(\omega) = \epsilon_\infty -
\omega_{ps}^2/ \omega^2$, so that the plasma frequency of the condensate is
$\omega_{ps,c}^2 = -\omega^2 \epsilon_{1,c}(\omega)$ in the limit of $\omega
\rightarrow 0$.  The frequency dependence of $\sqrt{-\omega^2 \epsilon_{1,c}
(\omega)}$ is shown in Fig.~\ref{fig:rho} for a variety of oxygen dopings for
$T\ll T_c$.  The low-frequency extrapolations employed in the Kramers-Kronig
analysis of the reflectance (typically below $40\,{\rm cm}^{-1}$) are included
to allow the $\omega \rightarrow 0$ values to be determined more easily.  The
estimate of $\omega_{ps,c}$ assumes that the response of $\epsilon_{1,c}
(\omega)$ at low frequency is dominated by the superconducting condensate.  The
overdoped material is known to have a large amount of low-frequency residual
conductivity for $T\ll T_c$, which may lead to an overestimate of
$\omega_{ps,c}$.  However, one of us (SVD) has developed a self-consistent
technique\cite{dordevic02} whereby $\epsilon_{2,c}(\omega)$ may be used to
calculate corrections to $\epsilon_{1,c}(\omega)$ and subsequently allow an
accurate determination of the value of $\omega_{ps,c}$.  The corrections are
typically small (less than a few \%); the values for $\omega_{ps,c}$ are listed
in Table~I, and are in good agreement with values recently obtained from
zero-field ESR studies.\cite{pereg04}

%
% Table I - c-axis parameters
%
\begin{table}[t]
%\vspace*{-0.3cm}%
\caption{The doping-dependent values in YBa$_2$Cu$_3$O$_{6+x}$ for the critical
temperature ($T_c$), and {\it c}-axis far-infrared conductivity ($\sigma_{dc}$)
measured just above $T_c$, strength of the condensate, expressed as a plasma
frequency ($\omega_{ps,c}$), and the penetration depth [$\lambda_c =
1/(2\pi\omega_{ps,c})$].}
\begin{ruledtabular}
\begin{tabular}{ccccc}
  $x$ &   $T_c$ (K) &  $\sigma_{dc}$ ($\Omega^{-1}$cm$^{-1}$)$^a$ &
  $\omega_{ps,c}$ (cm$^{-1}$) & $\lambda_c$ ($\mu$m) \\
  \cline{1-5}
  0.50  & 53   &   $9\pm 2$  &  $204\pm 20$ & 7.80  \\
  0.60  & 58   &  $12\pm 2$  &  $244\pm 20$ & 6.52  \\
  0.70  & 63   &  $14\pm 2$  &  $315\pm 30$ & 5.05  \\
  0.80  & 78   &  $27\pm 4$  &  $465\pm 35$ & 3.42  \\
  0.85  & 89   &  $47\pm 7$  &  $790\pm 50$ & 2.01  \\
  0.90  & 91.5 &  $88\pm 10$ & $1003\pm 60$ & 1.59  \\
  0.95  & 93.2 & $220\pm 20$ & $1580\pm 70$ & 1.01  \\
  0.99  & 90   & $450\pm 30$ & $2070\pm 90$ & 0.77  \\
\end{tabular}
\end{ruledtabular}
\footnotetext[1] {Taken at $\omega \rightarrow 0$ limit for $T\gtrsim T_c$.}%
\end{table}

%
% Figure 4
%
\begin{figure}[t]
%\vspace*{-0.7cm}
%
% eprint
%
\centerline{\includegraphics[width=3.2in]{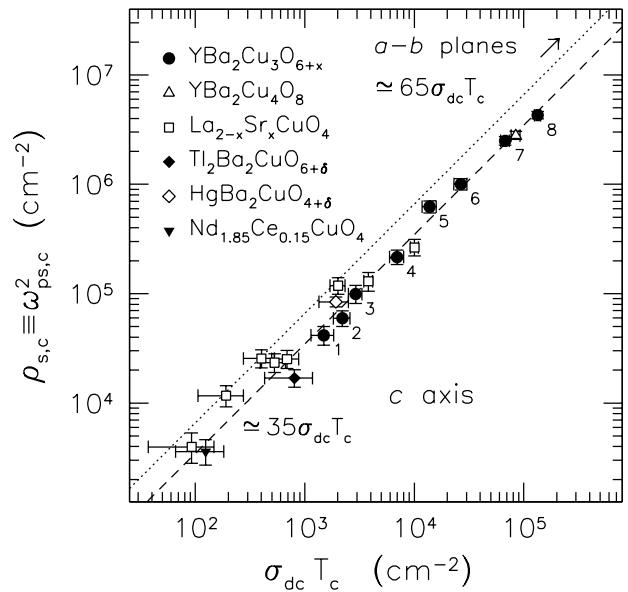}}%
%\vspace*{-0.1cm}%
%
% manuscript
%
%\centerline{\includegraphics[width=6.5in]{figure4.eps}}%
%
%\vspace*{-1.4cm}%
%
\caption{The optically-determined strength of the condensate $\rho_{s,c}$ vs
$\sigma_{dc}\,T_c$ along the $c$ axis in YBa$_2$Cu$_3$O$_{6+x}$ for various
oxygen dopings (labeled from the lowest to the highest oxygen doping in
ascending order), as well as results for other single and double-layer
copper-oxide superconductors. The data are described reasonably well by the
dashed line, $\rho_{s,c} \simeq 35\,\sigma_{dc}T_c$.  The dotted line is the
result due to Josephson coupling assuming a BCS isotropic gap in the
weak-coupling limit, $\rho_{s,c} \simeq 65\,\sigma_{dc}\,T_c$.  The arrow
indicates that the {\it a-b} plane data may also be scaled on the same dashed
line as the {\it c} axis data.\cite{homes04}}
%
%\vspace*{-0.3cm}%
\label{fig:lambda}
\end{figure}

%
% Always agrees with Josephson picture, more or less...
%
The optically-determined values for the superfluid density $\rho_{s,c}$ versus
$\sigma_{dc}\,T_c$ for YBCO are shown in the log-log plot in
Fig.~\ref{fig:lambda}. In addition, the {\it c}-axis results for
YBa$_2$Cu$_4$O$_8$ (Ref.~\onlinecite{basov94}), Tl$_2$Ba$_2$CuO$_{6+\delta}$
(Ref.~\onlinecite{basov99}), HgBa$_2$CuO$_{4+\delta}$
(Ref.~\onlinecite{homes04}), and La$_{2-x}$Sr$_x$CuO$_4$
(Ref.~\onlinecite{homes04}) are also shown. All the points fall on a line
approximated by the scaling relation $\rho_{s,c} \simeq 35\,\sigma_{dc}\,T_c$,
which is close to the result from Josephson coupling.\cite{homes04}  In the
log-log representation of Fig.~\ref{fig:lambda}, the numerical constant in the
scaling relation is the offset of the line.  The line may be shifted by
assuming different ratios between $2\Delta$ and $k_B\,T_c$; the initial value
of $\simeq 65$ was based on the weak-coupling value of $2\Delta/k_B\,T_c \simeq
3.5$, while the observed value of $\simeq 35$ implies a smaller ratio
$2\Delta/k_B\,T_c \simeq 2$.
Previous studies along the {\it c} axis of the cuprate materials considered the
dependence of $\rho_{s,c}$ with $\sigma_{dc}$ (Ref.~\onlinecite{dordevic02});
underdoped materials followed this scaling behavior reasonably well, but
deviations were observed for optimal and overdoped materials.
%
% Why does the BCS picture work so well?
%
It is surprising that the Josephson coupling result describes the scaling
behavior along the {\it c} axis as well as it does given the assumption of a
BCS {\it s}-wave isotropic energy gap, when there is strong evidence to suggest
that the energy gap in copper-oxygen planes of YBCO is {\it d}-wave in nature
and contains nodes.\cite{hardy93,ding96}  A possible explanation may be that
the {\it c} axis properties are particularly sensitive to the zone boundary
$(\pi,0),\,(0,\pi)$ part of the Fermi surface where the superconducting gap is
observed to open at $T_c$ in optimally-doped materials.\cite{chakravarty93,
xiang96,marel99,ioffe99,bernhard99,basov01} In this case the {\it d}-wave
nature of the superconducting gap is not probed and the assumption of an
isotropic gap is qualitatively correct, yielding a reasonable agreement between
theory and experiment.
%
% You can get Josephson coupling from more than an insulating weak link...
%
The Josephson result might have been expected for the underdoped materials
where the transport along the {\it c} axis was activated and considered
incoherent. However, it is less obvious for the optimally and overdoped
systems; the overdoped material in particular gave indications of anisotropic
3D normal-state transport, and as such some deviation from this behavior might
have been expected.  Thus, it would be tempting to assume that the observed
scaling $\rho_{s,c} \propto \sigma_{dc}\,T_c$ along the {\it c} axis justifies
the view that the coupling between the planes is always incoherent. However, it
is important to note that both tunnel junctions (SIS) in the case considered
here, as well as non-tunnel junctions (SNS) can show the Josephson effect with
a nearly identical Josephson current.\cite{kulik75,likharev79,golubov04}
Consequently, the qualitative agreement with Josephson coupling in the
optimally-doped and overdoped materials does not necessarily imply that the
normal-state transport is incoherent; that determination must be made from the
normal-state transport.

%
% This is the part for the conclusions.
%
\section{Conclusions}
The optical and transport properties in the normal state of YBCO suggest that
the material is showing signs of anisotropic 3D metallic transport at high
oxygen dopings for $T\simeq T_c$.  A scaling relation  $\rho_{s,c} \simeq
35\,\sigma_{dc}\,T_c$ is observed along the {\it c} axis, in agreement with the
result expected from Josephson coupling in the BCS weak limit case.  However,
the Josephson effect does not necessarily imply incoherent behavior between the
copper-oxygen planes, suggesting instead that the transport may indeed be
tending towards more coherent behavior in YBCO at higher oxygen dopings.

%
% Acknowledgements...
%
%\vspace*{-0.7cm}
\begin{acknowledgments}
%\vspace*{-0.4cm}%
We would like to thank D.~N.~Basov, S.~L.~Cooper, V.~J.~Emery, J.~C.~Irwin,
M.~V.~Klein, T.~R\~{o}\~{o}m, M.~Strongin, and J.~J.~Tu for helpful
discussions. This work was supported by the Natural Sciences and Engineering
Research Council of Canada, the Canadian Institute for Advanced Research, and
the Department of Energy under contract number DE-AC02-98CH10886.
\end{acknowledgments}
%\vspace*{-0.5cm}
%
%%%%%%%%%%%%%%%%%%%%%%%%%%%%%%%%%%%%%%%%%%%%%%%%%%%%%%%%%%%%%%%%%%%%%%%%%%%%%%
%
% References
%
\bibliography{trends}

\end{document}